# Traveling waves and localized structures:
## An alternative view of nonlinear evolution equations

Yair Zarmi

Jacob Blaustein Institutes for Desert Research

Ben-Gurion University of the Negev

Midreshet Ben-Gurion, 8499000, Israel

Given a nonlinear evolution equation in (1+$n$) dimensions, which has spatially extended traveling wave solutions, it can be extended into a system of two coupled equations, one of which generates the original traveling waves, and the other generates structures that are localized in the vicinity of the intersections of the traveling waves. This is achieved thanks to the observation that, as a direct consequence of the original evolution equation, a functional of its solution exists, which vanishes identically on the single-wave solution. This functional maps any multi-wave solution onto a structure that is confined to the vicinity of wave intersections. In the case of solitons in (1+1) dimensions, the structure is a collection of humps localized in the vicinity of soliton intersections. In higher space dimensions these structures move in space. For example, a two-front system in (1+3) dimensions is mapped onto an infinitely long and laterally bounded rod, which moves in a direction perpendicular to its longitudinal axis. The coupled systems corresponding to several known evolution equations in (1+1), (1+2) and (1+3) dimensions are reviewed.

## 1. Introduction

The interest in nonlinear systems that admit solutions in the form of spatially localized structures in (1+$n$) dimensions has been growing rapidly over the years. The usual approach focuses on finding finite-amplitude, spatially localized solutions of a single evolution equation [1-25]. Many works have considered the generation of localized structures in systems of coupled equations that are obtained as possible descriptions for a variety of physical systems (see. e.g., Refs. [26-43]).

This paper also focuses on the generation of localized structures in systems of coupled equations. The novelty is that the systems are simple extensions of nonlinear evolution equations in (1+$n$) dimensions, which have spatially extended traveling wave solutions such as solitons or fronts. (In this context, a soliton in (1+1) dimensions, while localized along the $x$-axis, is viewed as spatially extended in the $x$-$t$ plane.) The algorithm exploited leads to the extension of an evolution equation into a system of two coupled equations, one of which generates the traveling wave solutions of the original equation, whereas the second equation generates localized structures.

The starting point is a direct consequence of the evolution equation: Given the solution, $u$, of that equation, a functional, $R[u]$, exists, which vanishes identically when $u$ is a single-wave solution. The functional, $R[u]$, maps multi-wave solutions onto structure that are confined to the vicinity of wave intersections. In the case of solitons in (1+1) dimensions these structures are humps in the vicinity of soliton intersections. In the case of fronts in (1+1) dimensions of the Burgers equation these structures are KdV-like solitons on a half line. In more than one space dimension, the localized structures move in space. For example, the multi-soliton solutions of the Kadomtsev-Petviashvili (KP) II equation are mapped onto collections of humps that move in the $x$-$y$ plane, and a two-front solution of the Sine-Gordon equation in (1+3) dimensions is mapped onto an infinitely long, but laterally bounded, rod that moves in a direction perpendicular to its longitudinal axis.

A physically interesting outcome of this approach is that the emergence of solutions with more than one traveling wave (i.e., one soliton, or one front) may be viewed as the splitting of a single wave into several waves under the effect of the accompanying localized structure. Concurrently, generation of the localized structures may be viewed as a consequence of the interaction amongst spatially extended traveling waves.

The cases of several known evolution equations in (1+1), (1+2) and (1+3) dimensions are reviewed. The generation of localized structures from the multi-soliton solutions of evolution equations in (1+1) dimensions is reviewed in Section 2. In these cases, $R[u]$ describes humps in the $x$-$t$ plane. Section 3 is devoted to the Burgers equation. The images of its front solutions under $R[u]$ are KdV-like solitons on a half line. The generation of moving spatially localized structures in (1+2) dimensions in the case of the KP II equation is reviewed in Section 4. Section 5 reviews the generation of localized structures from multi-front solutions of the Sine-Gordon equation. In (1+1) dimensions, these are humps in the $x$-$t$ plane. In (1+2) dimensions, they are moving humps. In (1+3) dimensions, they are infinitely long straight rods, which move in directions that are perpendicular to their longitudinal axes. Concluding comments are presented in Section 6.

**2. Equations in (1+1) dimensions: Stationary localized structures**
In this Section, (1+1)-dimensional evolution equations that admit soliton solutions are discussed. The case of the KdV equation is analyzed in detail. Other equations, for which the results are similar to those of the KdV equation, are reviewed briefly. The NLS equation is discussed in detail, as finding $R[u]$ in that case is somewhat involved. Invariably, $R[u]$ maps multi-soliton solutions onto collections of humps localized in the $x$-$t$ plane in the vicinity of soliton intersections.

**2.1 The KdV equation**
The KdV equation,

$$u_t = 6uu_x + u_{xxx} \; , \tag{1}$$

is integrable. Its Lax pair is given by

$$\psi_{xx} = -(\lambda + u)\psi$$

$$\psi_t = \left(4\partial_x^3 + 6u\partial_x + 3u_x\right)\psi$$

(2)

### 2.1.1 The functional *R*[*u*]

Consider a single-wave solution,

$$u(t,x) = f(\xi) \quad , \quad (\xi = x + vt) \quad , \quad \left(f(\xi) \underset{|\xi| \to \infty}{\to} 0\right) .$$

(3)

Eq. (1) now becomes an ODE:

$$v f' = 6 f f' + f''' .$$

(4)

For *f*(*ξ*) to represent a single soliton, it must vanish as $\xi \to -\infty$. Exploiting this boundary condition, integration of Eq. (4) yields

$$v f = 3 f^2 + f'' .$$

(5)

Multiplying Eq. (4) by *f* and then integrating yields

$$\frac{1}{2} v f^2 = 2 f^3 + f f'' - \frac{1}{2}(f')^2 .$$

(6)

Eqs. (5) and (6) yield a result that is independent of the velocity, *v*:

$$f^3 + f f'' - (f')^2 = 0 .$$

(7)

Eq. (7) implies that the differential polynomial,

$$R[u] = u^3 + u u_{xx} - (u_x)^2 ,$$

(8)

vanishes identically when *u* is a single-traveling wave solution of Eq. (1) independently of its velocity, provided it vanishes for $\xi \to -\infty$. Furthermore, viewing

$$R[u] = u^3 + u u_{xx} - (u_x)^2 = 0$$

(9)

as an ODE in *x*, it is solved by the single-KdV soliton:

$$u = \frac{2k^2}{(\cosh[kx + \delta])^2} .$$

(10)

The time dependence of $u$ is determined by Eq. (1) to be given by $\delta = 4k^2 t + \delta_0$.

$R[u]$ maps a multi-soliton solution onto a collection of positive definite humps that are localized around soliton intersections and fall off exponentially away from their maxima. Figs. 1 and 2 present, respectively, a two-soliton solution and its image, $R[u]$. The two-soliton solution is given by:

$$u = 2\partial_x^2 \log\left\{1 + g_1 + g_2 + \left(\frac{k_1 - k_2}{k_1 + k_2}\right)^2 g_1 g_2\right\} \quad (11)$$

$$g_i = e^{2k_i(x + 4k_i^2 t + \delta_i)} \quad , \quad (i = 1,2)$$

### 2.1.2 Connection between $u$ and $R[u]$

Formal inversion of Eq. (8) yields for $u$ a non-local functional of $R[u]$. With $u$ expressed in terms of $R$, the Lax pair of Eqs. (2), becomes a Lax pair for $R(t,x)$. Such inversion may be executed in two ways. One way is to view Eq. (8) as an ODE in $x$ for $u(t,x)$, $t$ being a parameter, introduced through the time dependence of $R$. This equation yields solutions up to an arbitrary function of $t$:

$$u(t,x) = u_0(t,x) + F(t) \quad . \quad (12)$$

In Eq. (12), $u_0(t,x)$ may be chosen to vanish as $|t| \to \infty$. For soliton solutions, one has to impose

$$F(t) \equiv 0 \quad . \quad (13)$$

Another way to obtain $u[R]$ is to view Eq. (8) as a cubic equation for $u$. and adopt the real root:

$$u(t,x) = \frac{1}{3Q} - Q \quad \left(Q = \frac{(-(R + u_x) + \sqrt{(R + u_x)^2 + \frac{4}{27} u_{xx}^3}}{2}\right) \quad . \quad (14)$$

Employing Eq. (14) repeatedly yields $u$ as a functional of $R$ and all its spatial derivatives.

### 2.1.3 Extension to coupled system of solitons and humps

One may view $u$ and $R[u]$ as the solutions of a system of two coupled equations. Exploiting Eq. (8), Eq. (1) may be re-written in several forms, of which a simple one is:

$$u_t = \left\{u_x\left(2u^3 + (u_x)^2\right)/u^2\right\} + \left\{\partial_x(uR)/u^2\right\} \quad . \quad (15)$$

The part of Eq. (15) that does not contain coupling to $R[u]$,

$$u_t = \left\{ u_x \left( 2u^3 + (u_x)^2 \right) / u^2 \right\} , \qquad (16)$$

is obeyed by the single-soliton solution of Eq. (1), for which $R[u]$ vanishes. However, Eq. (16) is not obeyed by multi-soliton solutions of the KdV equation.

The dynamical equation for $R$ may have a variety of forms, depending on the specific procedure employed in obtaining it. Repeated application of Eq. (1) to $\partial_t R$, leads to the following equation:

$$\begin{aligned} R_t &= 6u R_x + R_{xxx} + 3 R_{KdV}[u] \\ \left( R_{KdV}[u] \right. &= \left. 6(u_x)^3 + u_x u_{xxxx} - u_{xx} u_{xxx} \right) \end{aligned} \qquad (17)$$

$R_{KdV}[u]$ vanishes on the single-soliton solution. Hence, it is localized around solitons intersections in multi-soliton solutions. (However, unlike $R[u]$, $R_{KdV}[u]$ is not positive definite.) Eqs. (15) and (17) are mutually consistent in the sense that, when $u$ is a single-soliton solution, Eq. (17) is solved by $R \equiv 0$ because $R_{KdV}[u]$ then vanishes.

Other equivalent forms of the equation of $R$ are obtained by employing the definition, Eq. (8), to express $u_{xx}$ in terms of $R$, leading to the identity:

$$\partial_x^n u = \partial_x^{n-2} \left( \frac{R + (u_x)^2 - u^3}{u} \right) , \quad (n \geq 2) . \qquad (18)$$

Exploiting Eq. (18), for $n = 3$ and 4 so as to eliminate $u_{xxx}$ and $u_{xxxx}$, respectively, in Eq. (17) yields for $R$ the equation:

$$R_t = 6u R_x + R_{xxx} + 3 \frac{u_x R_{xx} - u_{xx} R_x}{u} . \qquad (19)$$

If one exploits Eq. (18) for $n = 2$, so as to eliminate also $u_{xx}$, yet another equation for $R$ is obtained:

$$R_t = 6u R_x + R_{xxx} + 3 \left\{ R_x \left( u - \left( \frac{u_x}{u} \right)^2 \right) + R_{xx} \frac{u_x}{u} - \frac{R R_x}{u^2} \right\} . \qquad (20)$$

Both Eqs. (19) and (20) are solved by the trivial solution, $R \equiv 0$. For the latter, Eq. (15) becomes Eq. (16), which is solved by the single-KdV soliton solution, but not by multi-KdV soliton solutions.

Direct substitution shows that the system of coupled equations, comprised of Eq. (15) and any one of Eqs. (17), (19) or (20), is indeed solved by all multiple-soliton solutions of the KdV equation, Eq. (1), and by the corresponding hump structure generated by $R[u]$ of Eq. (8). To Solve for $R[u]$, one may either first solve Eq. (1) for $u(t,x)$ and then use Eq. (8) to construct $R[u]$, or solve one of Eqs. (17), (19) or (20) for $R[u]$ directly, for a given profile of $u(t,x)$.

Of the three combinations, the system comprised of Eqs. (15) and (17) offers new and interesting physical insight: In Eq. (15), $R[u]$ plays the role of a localized "obstacle", which leads to splitting of a single soliton into several solitons. Concurrently, the driving term, 3 $R_{KdV}[u]$ in Eq. (17), which represents the interaction amongst the intersecting solitons, is responsible for the generation of the positive definite hump, $R[u]$.

Finally, $R[u]$ of Eq. (8) and $R_{KdV}[u]$ of Eq. (17) are members of an infinite hierarchy of "special polynomials" – differential polynomials in $u$, which vanish when $u$ is a single-soliton solution. This point is briefly reviewed in Appendix I.

**2.2 Bi-directional KdV equation**
The functional form of the single-soliton solution of the Bi-directional KdV equation,

$$u_{tt} - u_{xx} - \left(6u u_x + u_{xxx}\right)_x = 0 \quad , \tag{21}$$

is the same as of a single-KdV-soliton, with a velocity, $v = (1 + 4 k^2)^{1/2}$, where $k$ is the wave number of the soliton [44]. Hence, $R[u]$ of Eq. (8) vanishes also on the single-soliton solution of Eq. (21). This can be also shown formally. For $u$ the single-wave form of Eq. (3) one obtains:

$$(v^2 - 1)f'' - (6ff')' - f'''' = 0 \quad . \tag{22}$$

Requiring that $f(\xi)$ vanish at $\xi \to -\infty$, integration of Eq. (22) once with respect to $\xi$ yields:

$$(v^2 - 1)f' - 3ff' - f''' = 0 \quad . \tag{23}$$

Multiplying Eq. (23) by $f(\xi)$ and integrating once, one obtains:

$$\frac{1}{2}(v^2 - 1)f^2 - 2f^3 - ff'' + \frac{1}{2}(f')^2 = 0 \quad . \tag{24}$$

Integrating Eq. (22) twice with respect to $\xi$ one obtains:

$$(v^2 - 1)f - 3f^2 - f'' = 0 \quad . \tag{25}$$

Eqs. (24) and (25) yield Eq. (7). Hence, $R[u]$ of (8) vanishes identically also on the single-soliton solution of Eq. (21).

An analysis similar to that leading to Eq. (15), leads to an alternative way of writing Eq. (21):

$$u_{tt} - u_{xx} - \left( \left\{ u_x \left( 2u^3 + (u_x)^2 \right) \big/ u^2 \right\} + \left\{ \partial_x (uR) \big/ u^2 \right\} \right)_x = 0 \quad . \tag{26}$$

The homogeneous part of Eq. (26),

$$u_{tt} - u_{xx} - \left( u_x \left( 2u^3 + (u_x)^2 \right) \big/ u^2 \right)_x = 0, \tag{27}$$

is solved only by the single-soliton solution of Eq. (21). The "obstacle" $R$ in Eq. (26) is responsible for the splitting into several-solitons.

Repeating steps similar to those applied in the case of the KdV equation completes the coupled system, yielding for $R$ the following equation:

$$\begin{gathered}
R_{tt} - R_{xx} - (6uR_x + R_{xxx})_x + R_{b-dir}[u] = 0 \\
\left( \begin{aligned} R_{b-dir}[u] = &(1+6u)\left(6u(u_x)^2 - 2(u_{xx})^2\right) + 60(u_x)^2 u_{xx} + 2(1+9u)u_x u_{xxx} \\ &- 2(u_{xxx})^2 - 2u_{xx} u_{xxxx} + 4u_x u_{xxxxx} - 6u(u_t)^2 + 2\left((u_{t,x})^2 - u_t u_{t,xx}\right) \end{aligned} \right)
\end{gathered} \quad . \tag{28}$$

$R_{b\_dir}[u]$ in Eq.(28) vanishes identically when $u$ is the single-soliton solution of Eq. (21). Hence, it is localized in the vicinity of soliton intersection regions when $u$ is a multi-soliton solution.

### 2.3 Sawada-Kotera equation
The single-soliton solution of the Sawada-Kotera equation,

$$u_t = 45u^2 u_x + 15uu_{xxx} + 15u_x u_{xx} + u_{xxxxx} , \qquad (29)$$

has the same functional form as the single-KdV soliton (with a velocity, $v = 16\ k^4$). Hence, $R[u]$ of Eq. (8) vanishes identically also on the single-Sawada-Kotera soliton. This may be seen in the manner employed previously, and also by writing Eq. (29) in the form:

$$u_t = 30u^2 u_x + 20uu_{xxx} + 10u_x u_{xx} + u_{xxxxx} + 5\partial_x R[u] , \qquad (30)$$

with $R$ given by Eq. (8). The equation,

$$u_t = 30u^2 u_x + 20uu_{xxx} + 10u_x u_{xx} + u_{xxxxx} , \qquad (31)$$

is a symmetry in the hierarchy of the KdV equation [45-54]. It is integrable and has the same infinite hierarchy of $N$-soliton solutions as the KdV equation for all $N \geq 1$, with a velocity, $v = 16\ k^4$ for each soliton [55]. $R[u]$ of Eq. (8) vanishes on the single-soliton solution of Eq. (31). Therefore, the driving term in Eq. (30), also vanishes on that same single-soliton solution. As a result, Eqs. (29) and (31) share the same single-soliton solution.

Steps similar to those employed in Section (2.1) yield an equation for the humps generated by $R$:

$$\begin{gathered} R_t = \left\{ 15\left(3Ru^2 + R_{xx}u + Ru_{xx}\right) + R_{xxxx} \right\}_x + R_{SK} \\ \begin{pmatrix} R_{SK} = 15\left(6u^4 + 12u(u_x)^2 + 9u^2 u_{xx} + 5(u_{xx})^2\right)u_x + \\ 15\left(u^3 + 7(u_x)^2 - uu_{xx}\right)u_{xxx} + 5\left(9uu_x - u_{xxx}\right)u_{xxxx} + 5u_x u_{xxxxx} \end{pmatrix} \end{gathered}. \qquad (32)$$

### 2.3 Modified KdV Equation
The single-soliton solution of the modified KdV equation,

$$u_t = 6u^2 u_x + u_{xxx} , \qquad (33)$$

is given by:

$$u(t,x) = \frac{k}{\cosh\{k(x+k^2 t + x_0)\}} \quad . \tag{34}$$

A procedure similar to the one used in the cases of Eqs. (1) and (17), shows that

$$R[u] = u^4 - (u_x)^2 + u u_{xx} \quad , \tag{35}$$

vanishes on the single-soliton solution, Eq. (34). When $u$ is a multi-soliton solution of Eq. (33), $R[u]$ generates positive definite humps in the vicinity of soliton intersections.

The corresponding coupled system of equations is:

$$u_t = \left\{ 3u^2 + \frac{u_{xx}}{u} + \partial_x \left(\frac{u_x}{u}\right) \right\} u_x + \partial_x \left(\frac{R}{u}\right) \quad , \tag{36}$$

$$\begin{aligned} R_t &= 6u^2 R_x + R_{xxx} + 3 R_{mKdV}[u] \\ \left(R_{mKdV}[u]\right. &\left.= 12 u u_x^3 - u_{xx} u_{xxx} + u_x u_{xxxx}\right) \end{aligned} \quad . \tag{37}$$

Here, again, the homogeneous part of Eq. (36) is obeyed by the single-mKdV-soliton solution, but not by its multi-soliton solutions, and $R_{mKdV}[u]$ vanishes when $u$ is a single-soliton solution.

**2.4 NLS Equation**
For the NLS equation,

$$u_t = i u_{xx} + 2i|u|^2 u \quad , \tag{38}$$

let us write the solution in the form

$$u(t,x) = e^{i\varphi(t,x)} \rho(t,x) \quad . \tag{39}$$

The real and imaginary parts of Eq. (38) then obey the equations:

$$2\rho^3 - \rho \varphi_x^2 + \rho_{xx} - \rho \varphi_t = 0 \quad , \tag{40}$$

$$2 \rho_x \varphi_x + \rho \varphi_{xx} + \rho_t = 0 \quad . \tag{41}$$

For a single-soliton solution, one writes

$$\rho = r(\xi) \quad, \quad \varphi = f(\eta) \quad, \quad \left( r(\xi) \underset{|\xi| \to \infty}{\to} 0 \right)$$
$$(\xi = k(x+vt) \quad, \quad \eta = q(x+\omega t))$$
(42)

Using Eq. (42), Eq. (41) yields for $r(\xi) \neq 0$

$$k(v + 2qf'(\eta))\frac{r'(\xi)}{r(\xi)} + q^2 f''(\eta) = 0. \quad (43)$$

There are two ways to analyze Eq. (43):

$$v + 2qf'(\eta) \neq 0 \Rightarrow \frac{r'(\xi)}{r(\xi)} = -\frac{q^2 f''(\eta)}{k(v + 2qf'(\eta))} = C, \quad (44)$$

where $C$ is some constant, or

$$f(\eta) = -\frac{v}{2q}\eta + f_0. \quad (45)$$

The choice of Eq. (44) does not allow for $r(\xi) \neq 0$, which vanishes as $|\xi| \to \infty$ - a prerequisite for a single-soliton solution. Hence, Eq. (45) is the only option.

Using Eqs. (41) and (42), Eqs. (40) yields

$$2r(\xi)^2 + k^2 \frac{r''(\xi)}{r(\xi)} = \omega q f'(\eta) + q^2 f'(\eta)^2 = D, \quad (46)$$

where $D$ is a constant. Rewriting Eq. (46) for $r(\xi)$ as

$$Q \equiv 2r(\xi)^3 + k^2 r''(\xi) - Dr(\xi) = 0, \quad (47)$$

and using the vanishing boundary condition at infinity, one obtains

$$Qr(\xi) - 2\int_{-\infty}^{\xi} Qr'(\xi)d\xi = r(\xi)^4 - k^2 r'(\xi)^2 + k^2 r(\xi)r''(\xi) = 0. \quad (48)$$

With Eq. (45) obeyed, Eq. (48) is equivalent to the requirement that the following functional of the solution of the NLS equation,

$$R[u, u^*] = u^3 u^* - u_x^2 + u u_{xx}, \quad (49)$$

vanishes identically when $u$ is a single-soliton solution:

$$u = k \frac{e^{i\gamma}}{\cosh \beta} \quad , \quad \left(\gamma = \omega t - qx + \varphi_0, \beta = k(x + vt + x_0), \omega = k^2 - v^2/4, q = v/2\right) . \quad (50)$$

(Direct substitution shows that the single-soliton solution, Eq. (50), obeys Eqs. (48) and (45).)

$R[u,u^*]$ maps multi-soliton solutions onto systems of humps that are localized in the vicinity of soliton intersections. Inversion of Eq. (48), and its complex conjugate yields $u$ and $u^*$ in terms of $R$ and $R^*$. Hence, the Lax pair for the NLS equation serves as a Lax pair for $R$ and $R^*$.

Figs 3 and 4 present the amplitudes of a two-soliton solution and of the resulting $R[u,u^*]$. The two-soliton solution is given by:

$$u = \frac{(f_1 + f_2)}{D}$$

$$f_1 = k_1 \frac{e^{i\gamma_1}}{\cosh \beta_1} \left\{ k_1^2 - k_2^2 - i k_2 (v_1 - v_2) \tanh \beta_2 + \frac{1}{4}(v_1 - v_2)^2 \right\} \quad , \quad f_2 = f_1|_{\{1 \leftrightarrow 2\}} \quad , \quad (51)$$

$$D = (k_1 - k_2)^2 + \left((v_1 - v_2)^2/4\right) + 2 k_1 k_2 \frac{\cosh(\beta_1 - \beta_2) - \cos(\gamma_1 - \gamma_2)}{\cosh \beta_1 \cos \beta_2}$$

where $\beta_n$ and $\gamma_n$ ($n = 1, 2$) are defined as in Eq. (50) for the wave numbers $k_1$ and $k_2$.

As in previous examples, the NLS equation is extended into a system of two coupled equations:

$$u_t = i\left(u^3 u^* + u_x^2\right)/u + i(R/u) . \quad (52)$$

The homogeneous part of Eq. (52),

$$u_t = i\left(u^3 u^* + u_x^2\right)/u , \quad (53)$$

is solved by the single-soliton solution, but not by the multi-soliton solutions of the NLS equation. The localized "obstacle", $R[u,u^*]$, leads to the splitting of a single soliton into several solitons. An equation for $R[u,u^*]$ is found to be:

$$\begin{gathered} R_t = 4 i u u^* R + i R_{xx} + 2 i R_{NLS} \\ \left(R_{NLS} = -u^2 u_x u_x^* - 3 u u^* u_x^2 + u^2 u^* u_{xx} + u_{xx}^2 - u_x u_{xxx}\right) \end{gathered} . \quad (54)$$

$R_{NLS}$, the driving term on the r.h.s. of Eq. (54), vanishes on the single-soliton solution of the NLS equation, and is localized around soliton intersections when $u$ is a multi-soliton solution.

### 3. Burgers Equation

Unlike the equations that have been discussed hitherto, the Burgers equation,

$$u_t = 2uu_x + u_{xx} , \tag{55}$$

is not integrable. It has an infinite family of front solutions [56]. The structure of the front solutions offers a different type of mapping by $R[u]$.

For a single-front solution,

$$u = f(\xi) , \quad \xi = x + vt . \tag{56}$$

Eq. (55) becomes an ODE:

$$vf' = 2ff' + f'' . \tag{57}$$

Let us now focus on a front with a vanishing boundary condition, so that $f(\xi)$ vanishes as $\xi \to -\infty$. Integration of Eq. (57) with respect to $\xi$ then yields

$$vf = f^2 + f' . \tag{58}$$

For the same boundary condition, simple manipulation on Eqs. (57) and (58) yields

$$f^2 f' + f f'' - (f')^2 = 0 . \tag{59}$$

Eq. (59) means that the differential polynomial

$$R[u] = u^2 u_x + u u_{xx} - (u_x)^2 , \tag{60}$$

vanishes identically when $u$ represents a single front, which vanishes as $x \to -\infty$ (or as $x \to +\infty$).

As in previous examples, Eq. (55) can be re-written as a sum of two terms:

$$u_t = uu_x + \left((u_x)^2/u\right) + \left(R[u]/u\right) . \tag{61}$$

The homogeneous part of Eq. (61),

$$u_t = u u_x + \left( (u_x)^2 / u \right) , \tag{62}$$

is obeyed by a single-front solution (with a vanishing boundary condition at either $+\infty$, or $-\infty$) of Eq. (55), but not by its multi-front solutions. The "obstacle" generated by $R[u]$ in Eq. (61) causes the splitting of the single front into several fronts.

The equation obeyed by $R[u]$ is:

$$\partial_t R[u] = 2 \partial_x (u R[u]) + \partial_x^2 R[u] - 2 \left( 2 u_x^3 - u_{xx}^2 + u_x u_{xxx} \right) . \tag{63}$$

The driving term on the r.h.s. of Eq. (63) vanishes when $u$ is a single-front solution with a vanishing boundary condition at either $+\infty$, or $-\infty$. When $u$ is a multi-front solution, $R[u]$ generates a structure that is confined to the vicinity of the lines of intersection of pairs of fronts that do not obey vanishing boundary conditions at infinity. For example, the two-front solution with a vanishing boundary condition for one front is given by:

$$u = \frac{k_1 e^{\xi_1} + k_2 e^{\xi_2}}{1 + e^{\xi_1} + e^{\xi_2}} \quad (\xi_i = k_i (x + k_i t)) . \tag{64}$$

$R[u]$ is then given by:

$$R[u] = \frac{k_1 k_2 (k_1 - k_2)^2 / 4}{\left( e^{-\frac{1}{2}(\xi_1 + \xi_2)} / 2 + \cosh \left[ \frac{1}{2} (\xi_1 - \xi_2) \right] \right)^2} . \tag{65}$$

Away from the front junction, ($|\xi_1|$ & $|\xi_2| \to \infty$, $|\xi_1 - \xi_2| = O(1)$), $R[u]$ tends to a KdV-like soliton along a half line:

$$R[u] \to \frac{k_1 k_2 (k_1 - k_2)^2 / 4}{\left( \cosh \left[ \frac{1}{2} (k_1 - k_2)(x + (k_1 + k_2) t) \right] \right)^2} . \tag{66}$$

A two-front solution and its image under $R[u]$ are presented in Figs. 5 and 6, respectively.

## 4. (1+2)-dimensional example: KP II equation

The single-soliton solution of the KP II equation,

$$\frac{\partial}{\partial x}\left(-4\frac{\partial u}{\partial t} + \frac{\partial^3 u}{\partial x^3} + 6u\frac{\partial u}{\partial x}\right) + 3\frac{\partial^2 u}{\partial y^2} = 0 \quad, \tag{67}$$

is also a single KdV-soliton [57]. As a result, an analysis similar to that performed in the case of the KdV equation, yields Eq. (8) as the form for $R[u]$. The dynamics of $R[u]$ have been discussed extensively [58], hence, the results are briefly reviewed and accompanied by exemplary figures. The construction of soliton solutions of the KP II equation is reviewed in Appendix II.

Exploiting Eq. (8), Eq. (67) may be re-written in the following form, which exposes the interaction between KP II soliton and the localized humps:

$$\frac{\partial}{\partial x}\left(-4\frac{\partial u}{\partial t} + \left\{u_x\left(2u^3 + (u_x)^2\right)\big/u^2\right\} + \left\{\partial_x(uR)\big/u^2\right\}\right) + 3\frac{\partial^2 u}{\partial y^2} = 0 \quad. \tag{68}$$

The second equation in the coupled system may be written as:

$$-4\frac{\partial R}{\partial t} + \frac{\partial^3 R}{\partial x^3} = \\ 6u_x u_{yy} - 4\left(3u^2 + u_{xx}\right)u_t + 6(u_x)^3 + \\ \partial_x\left(12uu_x + 3u^2 u_{xx} - 2(u_{xx})^2 + 2u_x u_{xxx} + uu_{xxxx}\right) \tag{69}$$

The driving term on the r.h.s. of Eq. (69) vanishes identically when $u$ is a single-soliton solution, hence is localized around soliton intersections when $u$ is a multi-soliton solution.

The image of a multi-soliton solution is a collection of positive definite humps around soliton intersections. These humps propagate in the *x-y* plane, imitating collisions of non-relativistic particles. Some of these collisions are elastic, some – are not. An example of a four-soliton solution of the KP II equation and the hump structure, onto which it is mapped by $R[u]$ of Eq. (8), are shown in Figs. 7 and 8, respectively for $t = +50$.

At positive times, the solution expands and the four "particles" of Fig. 8 move apart. At negative times, the solution shrinks, so that the four "particles" approach one another. At some finite time, they coalesce into one "particle". Thus, this system may be viewed as emulating the collision of non-relativistic particles through a single-particle intermediate state.

**5. Sine-Gordon equation**

The Sine-Gordon in (1+n) dimensions,

$$\partial_\mu \partial^\mu u + \sin u \equiv \left\{ \partial_t^2 - \left( \partial_{x_1}^2 + \ldots + \partial_{x_n}^2 \right) \right\} u + \sin u = 0, \quad \mu = 0,1,\ldots,n, \quad n = 1,2,3 \quad , \tag{71}$$

is integrable in (1+1) dimensions [59], but not in higher space dimensions [60-63]. Still, the algorithm presented by Hirota in the (1+1)-dimensional case [64] generates $N$-front (also called "kinks", "solitons" or "domain walls") solutions for any $N$ in the two- and three-dimensional versions of Eq. (71). Some constraints on solution parameters reflect the lack of integrability of the higher-dimensional equation [65, 66]). The algorithm for the construction of the front solutions is reviewed in Appendix IIII.

In all space dimensions, the functional,

$$R[u] = \frac{1}{2} \partial_\mu u \partial^\mu u + (1 - \cos u) \quad , \tag{72}$$

vanishes identically on the single-front solution [67, 68]. The second equation in the system of coupled equations (the equation for $R[u]$) is:

$$\partial_\mu \partial^\mu R = \partial_\mu \partial_\nu u \partial^\mu \partial^\nu u - \left( \partial_\mu \partial^\mu u \right)^2 \quad . \tag{73}$$

If one wishes to explicitly express the effect of the "obstacle" $R[u]$ on the front solutions, one may do so by applying an additional derivative to Eq. (71), leading to:

$$u_\nu \partial_\mu \partial^\mu u - \partial_\nu \left( \frac{1}{2} \partial_\mu u \partial^\mu u \right) + \partial_\nu R = 0 \quad . \tag{74}$$

In any space dimension, the homogeneous part of Eq. (74),

$$u_\nu \partial_\mu \partial^\mu u - \partial_\nu \left( \frac{1}{2} \partial_\mu u \partial^\mu u \right) = 0 \quad , \tag{75}$$

is solved for $u = h(p \cdot x)$, which depends only on one 4-momentum vector, $p$. This includes, of course, the single-front solution of Eq. (71). However, the multi-front solutions of Eq. (71) are not solutions of Eq. (75). The "obstacle" $R$ in Eq. (74) leads to the splitting of the solution into several fronts.

In (1+1) dimensions, $R[u]$ maps multi-front solutions onto positive definite humps that are localized in the $x$-$t$ plane. An example of a two-front solution of Eq. (71) in (1+1) dimensions and the image of that solution under $R[u]$ are presented in Figs. 9 and 10, respectively.

Interesting physics is revealed in higher space dimensions. As the properties of the localized structures have been discussed extensively in Refs. [67] and [68], only choice examples are presented here.

In (1+2) dimensions, if all pairs of momenta (see Appendix III) obey the constraint,

$$-1 \le p^{(i)}{}_\mu p^{(j)\mu} \le 1 \quad , \tag{76}$$

then the images of multi-front solutions under $R[u]$ describe positive definite humps, which propagate in the $x$-$y$ plane at velocities that are lower than $c = 1$, hence emulating free spatially extended relativistic particles, for which $R[u]$ plays the role of their mass density. An example of a two-front solution of Eq. (71) in (1+2) dimensions and the image of that solution under $R[u]$ are presented in Figs. 11 and 12, respectively.

In (1+3) dimensions, applying $R[u]$ to a two-front solution yields an infinitely long rod, which if Eq. (76) is obeyed, propagates at a velocity that is lower than $c = 1$ in a direction perpendicular to its longitudinal axis. A single rod image is presented in Fig. 13. Its lateral profile coincides with that of a (1+2)-dimensional hump, an example of which is shown in Fig. 12.

With more than two fronts, every triplet of the 4-momenta, from which the solution is constructed (see Eqs. (IIII.1)-(III.7)), must obey Eqs. (III.9) and (III.10). One way to achieve this is through Eq. (III.8). If all triplets of 4-mometa obey Eq. (8), then the resulting image, $R[u]$, of a multi-front solution describes positive definite profiles of parallel straight, infinitely long rods. The rods propagate at a velocity that is lower than $c = 1$ in a direction perpendicular to their longitudinal axis. An example of three parallel rods, onto which a three-front solution is mapped by $R[u]$ of Eq. (72), is presented in Fig. 14. If, however, Eq. (III.8) is not obeyed by all triplets of 4-momenta, then $R[u]$ maps multi-front solutions onto systems of intersecting rods. Each triplet propagates at the speed of light ($c = 1$). An example of three intersecting rods, onto which a three-front solution is mapped by $R[u]$, is presented in Fig. 15.

## 6. Concluding comments

In this paper it has been shown that an evolution equation in $(1+n)$ dimensions, which has spatially extended single- and multiple-wave solutions, can be extended into a system of two equations, which represent coupling between the traveling wave solutions of the original equation and structures that are localized in the vicinity of wave intersections.

The procedure proposed here ensures that the resulting system of two coupled equations is solved by the traveling wave solutions of the original evolution equation, and, concurrently, by the localized structures generated by the functional $R[u]$. However, two open questions regarding the system of two coupled equations call for future research:

1. The integrability of the system;

2. Finding whether the system has other solutions beyond the ones discussed in the paper.

## Appendix I: A hierarchy of "special ploynomials"

There is an infinite hierarchy of differential polynomials of the solution of Eq. (1), which vanish on the single-soliton solution. Some are localized around soliton intersections - others are not. A systematic way to construct the hierarchy exploits the notion scaling weight [45]. If one assigns the triplet $\{\partial_t, \partial_x, u\}$ the scaling weights of $\{3,1,2\}$, respectively, then each term in Eq. (1) has scaling weight 5. Writing the most general differential polynomial with a given scaling weight, $W$, and finding the coefficients for which that polynomial vanishes identically on a single soliton, one finds the members of the infinite hierarchy. in the case of the KdV equation, the lowest scaling-weight, for which a differential polynomial that vanishes identically on the single-soliton solution exists, is $W = 3$. There are two non-local polynomials, given by

$$R^{(3,1)} = u_x + \partial_x^{-1} u \quad , \quad R^{(3,2)} = 3u_x + 3\partial_x^{-1} u - \left(\partial_x^{-1} u\right)^3$$

$$\left(\partial_x^{-1} F \equiv \frac{1}{2}\left(\int_{-\infty}^{x} F\,dx - \int_{x}^{+\infty} F\,dx\right)\right) \tag{I.1}$$

The number of these "special polynomials" grows rapidly with $W$. A local polynomial appears for the first time at $W = 6$; it is $R[u]$ of Eq. (8).

**Appendix II: Construction of front solutions of KP II equation**
The line-soliton solutions of Eq. (3) are obtained through a Hirota transformation [57]:

$$u(t,x,y) = 2\partial_x^2 \log\{f(t,x,y)\} \ . \tag{II.1}$$

The function $f(t,x,y)$ is given by

$$f(t,x,y) \equiv f\left(M,N;\vec{\xi}\right) =$$

$$\begin{cases} \sum_{i=1}^{M} \xi_M(i)\exp\left(\theta_i(t,x,y)\right) & N=1 \\ \sum_{i=1}^{M} \xi_M(i)\exp\left(\sum_{j=1, j\neq i}^{M} \theta_j(t,x,y)\right) & N=M-1 \\ \sum_{1\leq i_1<...<i_N\leq M} \xi_M(i_1,...,i_N)\left(\prod_{1\leq j<l\leq N}\left(k_{i_l}-k_{i_j}\right)\right)\exp\left(\sum_{j=1}^{N}\theta_{i_j}(t,x,y)\right) & 2\leq N\leq M-2 \end{cases} , \tag{II.2}$$

$$k_1 < k_2 < ... < k_M \ , \tag{II.3}$$

$$\theta_i(t,x,y) = -k_i x + k_i^2 y - k_i^3 t \ . \tag{II.4}$$

In Eq. (II.2), $M$ is the size of a set of wave numbers. Each set of $N$ wave numbers is one of the $\binom{M}{N}$ subsets of $\{k_1,\ldots,k_M\}$. In the following, $f(t,x,y)$ and $u(t,x,y)$ will be denoted, respectively, by $f(M,N;\vec{\xi})$ and $u(M,N;\vec{\xi})$.

To exclude singular solutions of Eq. (II.1), one requires

$$\xi_M(i_1,\ldots,i_N) \geq 0 \ . \tag{II.5}$$

Apart from positivity, the coefficients, $\xi_M(i)$, with $N=1$ and $N=M-1$, may assume arbitrary values. However, for $2 \leq N \leq M-2$, $\xi_M(i_1,\ldots i_N)$ are constrained by the Plücker relations (see, E.g. [69]). For example, for $(M,N) = (4,2)$ one finds a single Plücker relation:

$$\xi_4(1,2)\xi_4(3,4) - \xi_4(1,3)\xi_4(2,4) + \xi_4(1,4)\xi_4(2,3) = 0 \ . \tag{II.6}$$

**Appendix III: Construction of front solutions of Sine-Gordon equation**

An $N$-front solution of Eq. (54) is constructed in terms of two auxiliary functions [64]:

$$u(x;P) = 4\tan^{-1}\left[g(x;P)/f(x;P)\right] \ , \tag{III.1}$$

$$P \equiv \left\{p^{(1)}, p^{(2)}, \ldots, p^{(N)}\right\} \ . \tag{III.2}$$

In Eqs. (III.1) and (III.2), $x$ is the coordinate 4-vetor, and $p^{(i)}$, $i=1, 2,..N$, are momentum 4-vector parameters. The functions $g(x;P)$ and $f(x;P)$ are given by:

$$g(x;P) = \sum_{\substack{1 \leq n \leq N \\ n\ \text{odd}}} \left( \sum_{1 \leq i_1 < \cdots < i_n \leq N} \left\{ \prod_{j=1}^{n} \varphi(x;p^{(i_j)}) \prod_{i_l < i_m} V(p^{(i_l)},p^{(i_m)}) \right\} \right) , \tag{III.3}$$

$$f(x;P) = 1 + \sum_{\substack{2 \leq n \leq N \\ n\ \text{even}}} \left( \sum_{1 \leq i_1 < \cdots < i_n \leq N} \left\{ \prod_{j=1}^{n} \varphi(x;p^{(i_j)}) \prod_{i_l < i_m} V(p^{(i_l)},p^{(i_m)}) \right\} \right) , \tag{III.4}$$

where

$$\varphi(x;p^{(i)}) = e^{\xi_i + \delta_i} \ , \quad \left(\xi_i = p^{(i)}_\mu x^\mu\right) \ , \tag{III.5}$$

$$p^{(i)}_\mu p^{(i)\,\mu} = -1 \ . \tag{III.6}$$

In Eq. (III.5), $\delta_i$ are constant arbitrary phases. Finally,

$$V(p,p') = \frac{1 + p_\mu p'^\mu}{1 - p_\mu p'^\mu} \quad . \tag{III.7}$$

The lack of integrability of the Sine-Gordon equation beyond (1+1) dimensions [60-63] shows up through constraints on the parameter vectors, $p^{(i)}$, in $N$-front solutions for all $N \geq 3$ [65, 66].

In (1+2), dimensions the constraint on the momentum vectors is that only two of them, $p^{(1)}$, and $p^{(2)}$, are linearly independent [65]. All other momenta must be linear combinations of these two:

$$p^{(i)} = \alpha_i p^{(1)} + \beta_i p^{(2)} \quad , \quad (i > 2) \quad . \tag{III.8}$$

In (1+3)-dimensions, the components of each triplet of vectors (total of $\binom{N}{3}$ triplets) must obey the constraint [65]:

$$(\Delta_0)^2 = (\Delta_x)^2 + (\Delta_y)^2 + (\Delta_z)^2 \quad . \tag{III.9}$$

Denoting by $p^{(i)}$, $p^{(j)}$, and $p^{(k)}$ ($1 \leq i \neq j \neq k \leq N$) the three vectors in a triplet, $\Delta_0$, $\Delta_x$, $\Delta_y$ and $\Delta_z$ are defined by:

$$\Delta_x = \begin{vmatrix} p_0^{(i)} & p_y^{(i)} & p_z^{(i)} \\ p_0^{(j)} & p_y^{(j)} & p_z^{(j)} \\ p_0^{(k)} & p_y^{(k)} & p_z^{(k)} \end{vmatrix} , \quad \Delta_y = \begin{vmatrix} p_0^{(i)} & p_z^{(i)} & p_x^{(i)} \\ p_0^{(j)} & p_z^{(j)} & p_x^{(j)} \\ p_0^{(k)} & p_z^{(k)} & p_x^{(k)} \end{vmatrix} , \quad \Delta_z = \begin{vmatrix} p_0^{(i)} & p_x^{(i)} & p_y^{(i)} \\ p_0^{(j)} & p_x^{(j)} & p_y^{(j)} \\ p_0^{(k)} & p_x^{(k)} & p_y^{(k)} \end{vmatrix}$$

$$, \tag{III.10}$$

$$\Delta_0 = \begin{vmatrix} p_x^{(i)} & p_y^{(i)} & p_z^{(i)} \\ p_x^{(j)} & p_y^{(j)} & p_z^{(j)} \\ p_x^{(k)} & p_y^{(k)} & p_z^{(k)} \end{vmatrix}$$

References
1. B.B. Kadomtsev and V.I. Petviashvili, *Sov. Phys. Dokl.* **15**, 539–541 (1970).
2. G. 't Hooft, *Nucl. Phys.* B **79**, 276-284 (1974).
3. A.M. Polyakov, *JETP Lett.* **20**, 194-195 (1974).
4. A. Davey and K. Stewartson, *Proc. R. Soc. A* **338**, 101–110 (1974).
5. R.F. Dashen, B. Hasslacher and A. Neveu, *Phys. Rev.* D **10**, 4114-4129 (1974).
6. C.G. Callan and D.J. Gross, *Nucl. Phys.* B **93**, 29-55 (1975).
7. S. Coleman, *Phys. Rev.* D **11**, 2088-2097 (1975).
8. S. Mandelstam *Phys. Rev.* D **11**, 3026-3030 (1975).
9. L.D. Faddeev and V.E. Korepin, *Phys. Rep.* **42**, 1-87 (1978).
10. D.J. Kaup and A.C. Newell, *Proc. Roy. Soc. Lond.* A **361**, 413-446 (1978).
11. E. Witten, *Nucl. Phys.* B **160**, 57-115 (1979).
12. L.P. Nizhnik, *Dokl. Akad. Nauk.* **254**, 332-335 (1980);
13. O.J.P Eboli and G.C. Marques, *Phys. Rev.* B **28**, 689-696 (1983).
14. B.G. Konopelchenko and V.G. Dubrovsky, *Phys. Lett.* A **102**, 15-17 (1984).
15. A.P. Veselov and S.P. Novikov, *Dokl. Akad. Nauk. SSSR*, **279**, 20-24 (1984);
16. B.G. Konopelchenko, *Inverse Problems*, **7**, 739-753 (1991).
17. R. Camassa and D.D Holm, *Phys. Rev. Lett.* **71**, 1661–1664 (1993).
18. P. Rosenau and J.M. Hyman JM. *Phys. Rev. Lett.* **70**, 564-567 (1993).
19. M. Boiti, L. Martina and F. Pempinelli, *Chaos Solitons Fractals* **5** 2377–2417 (1995).
20. M.J. Ablowitz, S. Chakravarty, A.D. Trubatch and J. Villaroel, *Phys. Lett A* **267**, 132-146 (2000).
21. S.N.M. Ruijsenaars, NATO SCIENCE SERIES, SERIES II: MATHEMATICS, PHYSICS AND CHEMISTRY vol. 35, pp. 273-292 (2001).
22. J. Maldacena, G. Moorec and N. Seiberg, *J. High Energy Phys.* **27**, 062 (2001).
23. X.-y. Tang, S.-y. .Lou and Y. Zhang, *Phys. Rev. E* **66**, 046601 (2002).
24. G.I. Burde, *Phys. Rev. E* **84**, 026615 (2011).
25. N. Manton and P. Sutcliffe, *Topological fronts*. Cambridge University Press; 2007.
26. E. Golobin, B.A. Malomed and A.V Ustinov, *Phys. Lett. A* **266**, 67-75 (2000).
27. S. Nakagiri and J. Ha, *Taiwanese J. Math.* **5**, 297-315 (2001).
28. K.R. Khusnutdinova and D.E. Pelinovsky, *Wave Motion* **38**: 1-10 (2003).
29. S.D. Griffiths, R.H.J. Grimshaw and K.R. Khusnutdinova, *Th. Math. Phys*. **137**, 1448-1458 (2003).
30. J. Zhang, *Math. Meth. In the Appl. Sci.* **26**, 11-25 (2003).

31. T. Alagesan, Y. Chung and K. Nakkeeran, *Chaos, Fronts and Fractals* **21**, 879-882 (2004).
32. V. Vasumathi and M. Daniel, *Phys. Lett. A* **373**, 76-82 (2008).
33. A. Gröndlund, B. Eliasson and M. Marklund, *EPL*, **86**, 24001 (2009).
34. A. Salas, *Nonlin. Anal.: Real World Appl.* **11**, 3930-3935 (2010).
35. D. Shang, *Appl. Math. & Comp.* **217**, 1577-1583 (2010).
36. Y.M. Zhao, H. Liu and Y. Yang, *J. Appl. Math. Sci.* **5**, 1621-1629 (2011).
37. A. Ankiewicz, N. Devine, M. Ünal, A. Chowdury and N. Akhmediev, *J. Opt.* **15**, 064008 (2013).
38. Y.M. Zhao, *J. Appl. Math.* **2014**, 534346 (2014).
39. G. Mu, Z. Qin and R. Grimshaw, *SIAM J. Appl. Math.* **75**, 1-20 (2105).
40. A. Degasperis, S. Wabnitz and A.B. Aceves, *Phys. Lett. A* **379**, 1067-1070 (2015).
41. W.P. Zhong, M. Belić and B.A. Malomed *Phys. Rev E*, **92**, 053201 (2015).
42. E. Yomba and G.A. Zaken, *Chaos*, **26**, 0883115 (2016).
43. Y.Y Wang, C.Q. Dai, G.Q. Zhou, Y. Fan and L. Chen, *Nonlin. Dyn.* **87**, 67-73 (2017).
44. R. Hirota, *J. Math. Phys.*, **14**, 810-814 (1973).
45. C.S. Gardner, J.M. Greene, M.D. Kruskal and M. Miura, *Comm. Pure Appl. Math.* **27**, 97-133 (1974).
46. M.J. Ablowitz, D.J. Kaup, A.C Newell and H. Segur, *Stud. Appl. Math.* **53**, 249-315 (1974).
47. V.E. Zakharov and S.V. Manakov, *Sov. Phys. JETP* **44**, 106-112 (1976).
48. M.J. Ablowitz and H. Segur, *Solitons and the Inverse Scattering Transforms* (SIAM, Philadelphia, 1981).
49. S.P. Novikov, S.V. Manakov, L.P. Pitaevskii and V.E. Zakharov, *Theory of Solitons*, (Consultant Bureau, New York, 1984).
50. A.C. Newell, *Solitons in Mathematics and Physics*, (SIAM, Philadelphia, PA, 1985).
51. P.J. Olver, *Applications of Lie Groups to Differential Equations* (Springer-Verlag, New York, 1986).
52. Y. Kodama, Normal Form and Solitons, pp. 319-340 in *Topics in Soliton Theory and Exactly Solvable Nonlinear Equation*, ed. by M.J. Ablowitz et al. (World Scientific, Singapore, 1987).
53. M.J. Ablowitz and P.A. Clarkson, *Solitons, Nonlinear Evolution Equations and Inverse Scattering* (Cambridge University Press, 1991).
54. Y. Hiraoka and Y. Kodama, Normal Form and Solitons, Lecture notes, Euro Summer School 2001, The Isaac Newton Institute, Cambridge, August 13-24 (2001).

55. K. Sawada and T. Kotera, *Progr. Th. Phys.* **51**, 1355-1367 (1974).
56. G.B. Whitham, *Linear and Nonlinear Waves* (Wiley, New York, 1974).
57. R. Hirota, *Physica D* **18**, 161-170 (1986).
58. Y. Zarmi, *Nonlinearity*, **27**, 1499-1523 (2014).
59. M.J. Ablowitz and P.A. Clarkson PA. *Nonlinear Evolution Equations and Inverse Scattering* London Mathematical Society, 1991.
60. M.J. Ablowitz, A. Ramani and H. Segur, *J. Math. Phys.* **21**, 715-721 (1980).
61. J. Weiss, M. Tabor and G. Carnevale, *J. Math. Phys.* **24**, 522-526 (1983).
62. J. Weiss, *J. Math. Phys.* **25**, 2226-2235 (1984).
63. P.A. Clarkson, *Lett. Math. Phys.* **10**, 297-299 (1985).
64. R. Hirota, *J. Phys. Soc. Japan* **33**, 1459-1963 (1972).
65. R. Hirota, *J. Phys. Soc. Japan* **35**, 1566 (1973).
66. K.K Kobayashi and M. Izutsu, *J. Phys. Soc. Japan* **41**, 1091-1092 (1976).
67. Y. Zarmi, *PloS ONE* **113**, 0148993 (2016).
68. Y. Zarmi, *PloS ONE* **137**, 0175783 (2017).
69. R. Hirota, *The Direct Method in Soliton Theory*, Cambridge University Press, (2004).

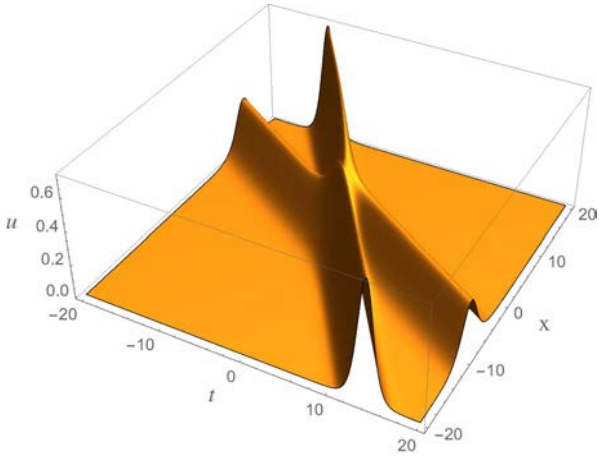

Fig. 1 Two-soliton solution of KdV equation. Wave numbers: $k_1 = 0.4$, $k_2 = 0.6$.

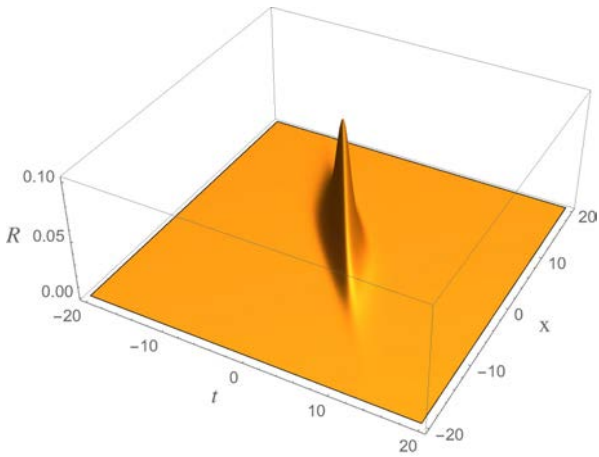

Fig. 2 $R[u]$ of Eq. (8) for KdV two-soliton solution of Fig. 1.

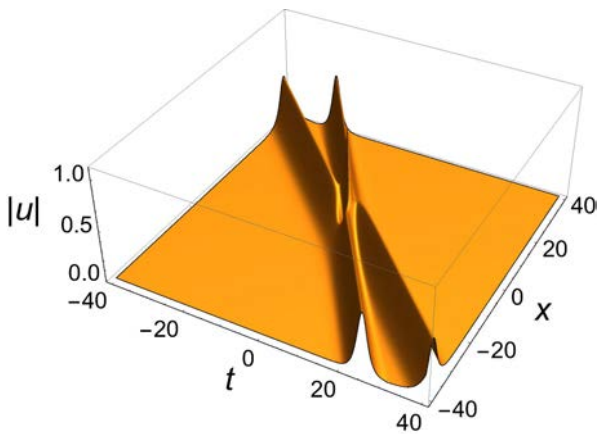

Fig. 3 Amplitude of two-NLS-soliton solution (Eq. (51)). $k_1 = .5$; $k_2 = .6$; $v_1 = 1$; $v_2 = 1.5$.

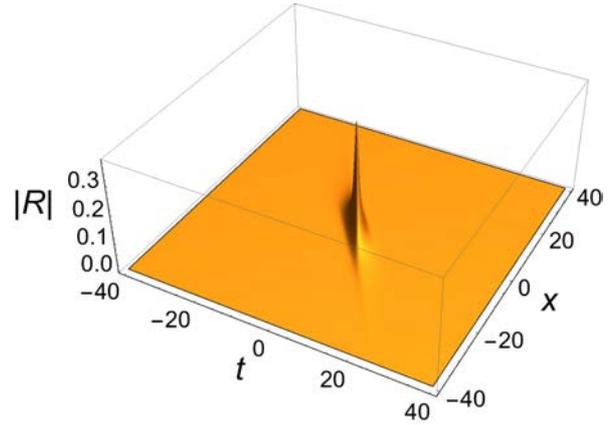

Fig. 4 Amplitude of $R[u,u^*]$ (Eq. (49)) for two-NLS-soliton solution of Fig. 3.

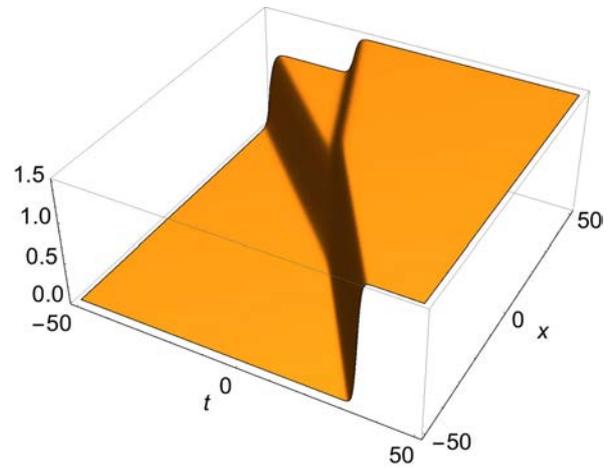

Fig. 5 Two-front solution of Burgers equation (Eq. (64)). Wave numbers: $k_1 = 1.$, $k_2 = 1.5$.

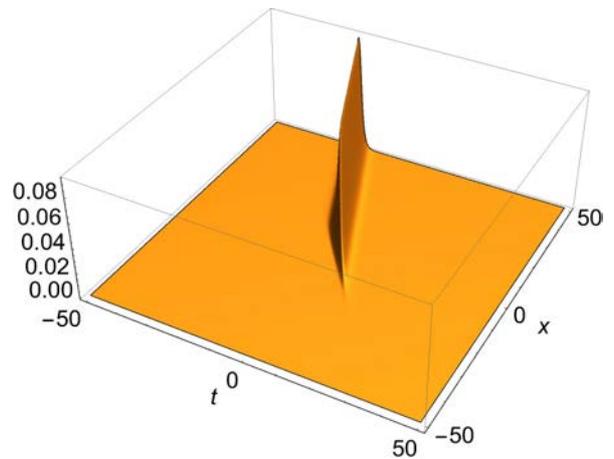

Fig. 6 $R[u]$ (Eq. (60)) for two-Burgers front solution of Fig. 5.

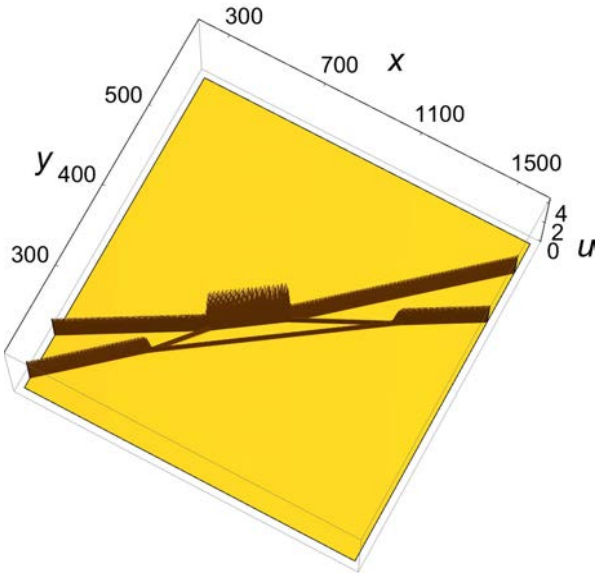

Fig. 7 Four-KP II-soliton solution (see Appendix II). Wave numbers:
$k_1 = 1., k_2 = 2., k_3 = 3., k_4 = 4.$
$\xi_{ij} = 2\dfrac{(j-i)}{(i+2)(j+2)}$. $t = +50$.

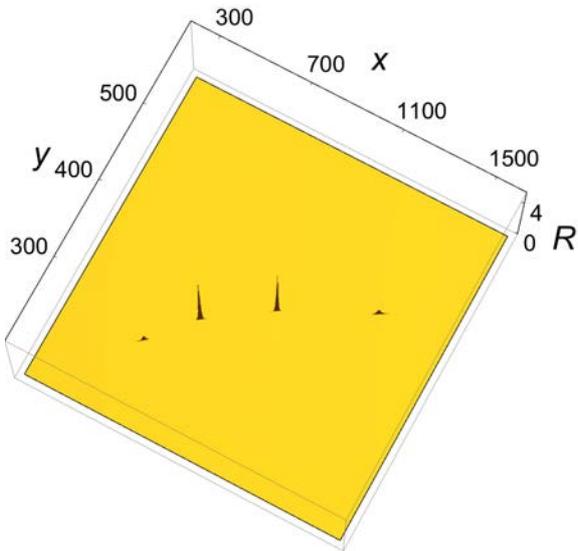

Fig. 8 $R[u]$ (Eq. (8)) for four-soliton solution of Fig. 7.

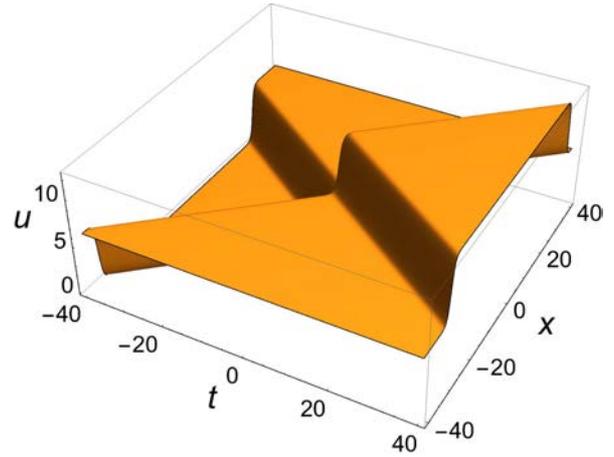

Fig. 9 Two-front solution of (1+1)-dimensional Sine-Gordon equation (see Appendix III). Parameter vectors: $p^{(1)} = \{1, -\sqrt{2}\}, p^{(2)} = \{2, \sqrt{5}\}, \delta_1 = \delta_2 = 0$.

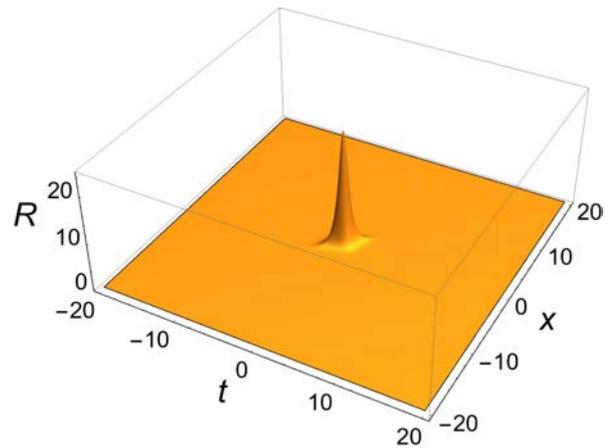

Fig. 10 $R[u]$ (Eq. (72)) for two-front solution of Fig. 9.

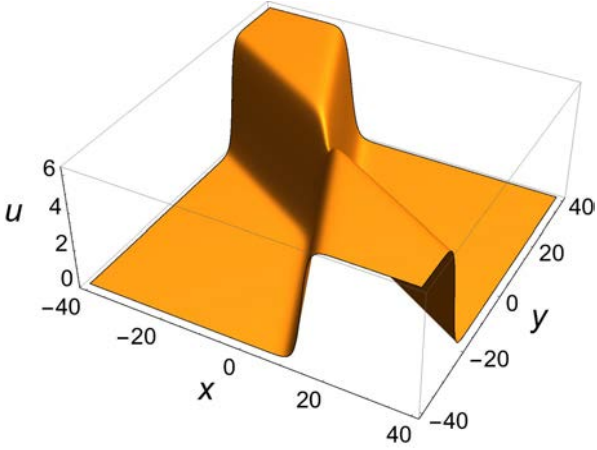

Fig. 11 Two-front solution of (1+2)-dimensional Sine-Gordon equation (see Appendix III) in its rest frame. Parameter vectors:

$$p^{(1)} = \{0, \cos(\pi/8), \sin(\pi/8)\}$$
$$p^{(2)} = \{0, \cos(\pi/3), \sin(\pi/3)\}.$$

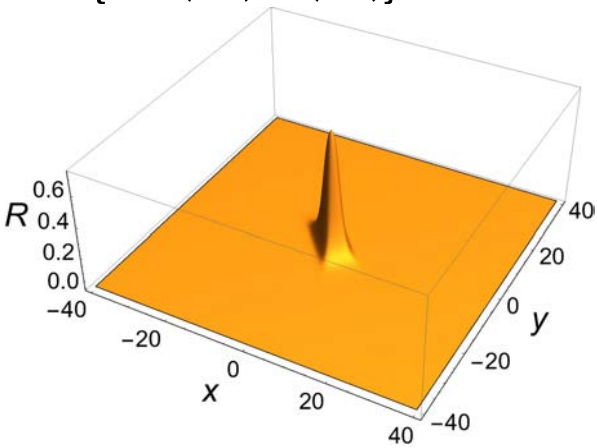

Fig. 12 $R[u]$ (Eq. (72)) for two-front solution of Fig. 11.

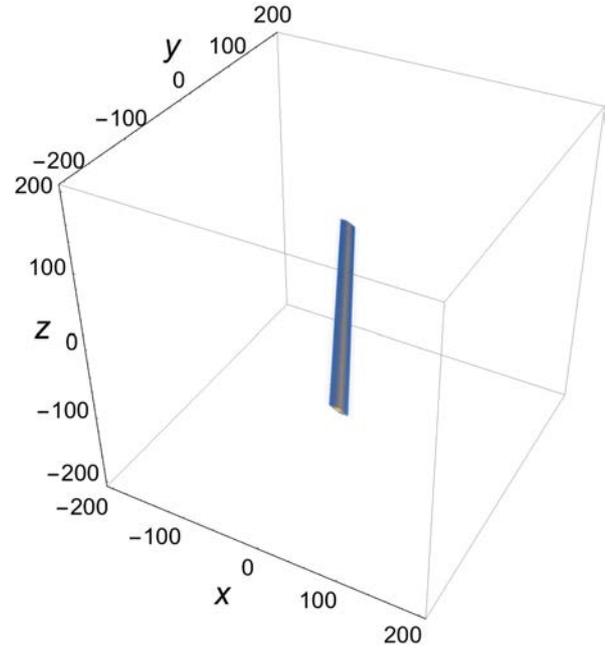

Fig. 13 Single-rod image (Eq. (72)) of two-front solution of Sine-Gordon equation in (1+3) dimensions (see Appendix III). Parameter vectors:

$$p^{(1)} = \{1, \cos(\pi/5), \sin(\pi/5, 1)\}$$
$$p^{(2)} = \{2, \cos(\pi/4), \sin(\pi/4), 2\}.$$
$$\delta_1 = \delta_2 = 0$$

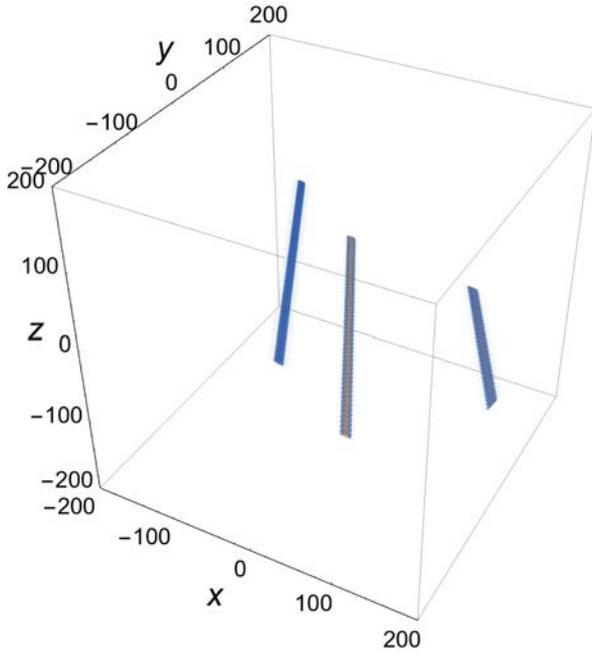 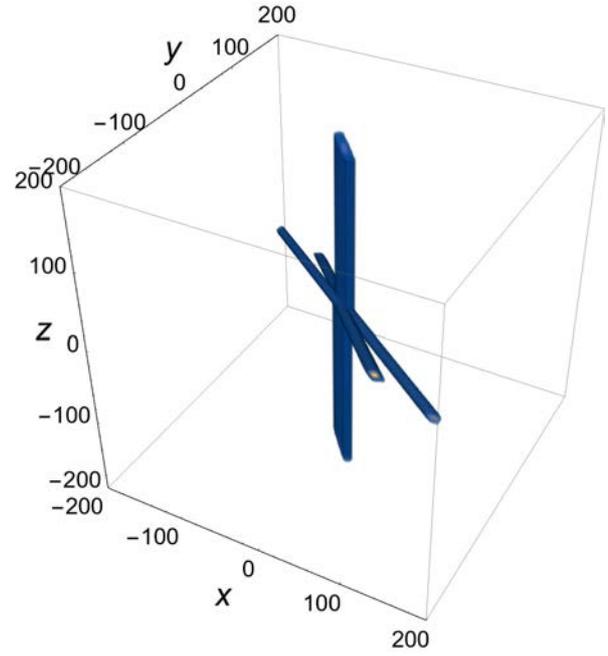

Fig. 14 Three parallel rod image (Eq. (72)) of three-front solution of Sine-Gordon equation in (1+3) dimensions (see Appendix III). Parameter vectors:

$p^{(1)} = \{1, \cos(\pi/5), \sin(\pi/5, 1)\}$
$p^{(2)} = \{2, \cos(\pi/4), \sin(\pi/4), 2\}$
$p^{(3)} = \alpha_1 p^{(1)} + \alpha_2 p^{(2)}$
$\alpha_1 = -1.65449, \alpha_2 = 2.60004$
$\delta_1 = 0., \delta_2 = 50., \delta_3 = -50$

Fig. 15 Three intersecting rod image (Eq. (72)) of three-front solution of Sine-Gordon equation in (1+3) dimensions (see Appendix III). Parameter vectors:

$p^{(1)} = \{0, \cos(\pi/5), \sin(\pi/5, 0)\}$
$p^{(2)} = \{0, \cos(\pi/4), \sin(\pi/4), 0\}$
$p^{(3)} = \{1, \cos(\pi/3), \sin(\pi/3), 1\}$
$\delta_1 = \delta_2 = \delta_3 = 0.$